\documentstyle[prl,aps,epsfig,preprint]{revtex}
\newcommand{\beq}{\begin{equation}}
\newcommand{\eeq}{\end{equation}}
\newcommand{\beqa}{\begin{eqnarray}}
\newcommand{\eeqa}{\end{eqnarray}}

\tightenlines 

\begin{document}

\title{\bf The Charge Ordered State from Weak to Strong Coupling}

\vskip 2cm
\author{S. Ciuchi$^{\dagger}$,
F. de Pasquale$^{\P}$,}

\address{{$^\dagger$} Dipartimento di Fisica and INFM UdR l'Aquila, 
Universit\`a de
L'Aquila, via Vetoio, I-67100 Coppito-L'Aquila, Italy}
\address{$^{\P}$ Dipartimento di Fisica and INFM UdR Roma 1,
Universit\`a di Roma "La Sapienza",
piazzale Aldo Moro 5, I-00185 Roma, Italy}

\maketitle

\vskip 0.5cm
\begin{abstract}
We apply the Dynamical Mean Field Theory to the problem of charge ordering. 
In the normal state as well as in the Charge Ordered (CO) state
the existence of polarons, i.e. electrons strongly coupled to local lattice
deformation,
is associated to the qualitative properties of the
Lattice Polarization Distribution Function (LPDF).
At intermediate and strong coupling 
a CO state characterized by a certain amount of thermally activated 
defects arise from the
spatial ordering of preexisting randomly distributed polarons. 
Properties of this particular CO state gives a qualitative understanding
of the low frequency behavior of optical conductivity of $Ni$
perovskites.
\end{abstract}

\narrowtext

\vskip 30pt
{\large {\bf 1. INTRODUCTION}}
\vskip 15pt
Recently there has been a renewed interest for the charge ordering 
transition which has been found 
associated to lattice displacements in Cuprates
Nickelates and Manganites.

Charge stripes order has been detected in Neodimium doped Cuprates  
($La_{1.475}Nd_{0.4}Sr_{0.125}CuO_4$)\cite{LaNd}
and conjectured in $LASCO$\cite{LASCO}. X ray studies of
$BISCO$ also shown a modulated structure of $CuO$
planes \cite{BISCO}. 
Commensurate
charge order appears in doped Nickelates \cite{LASCO} also related with 
peculiar magnetic properties\cite{Ni}, and finally
large lattice distortions have been found in Manganites 
which can be associated to either commensurate or incommensurate
charge ordering \cite{Mn}.
The amplitude of such a distortion increases from Cuprates to Manganites,
a fact that may support the hypothesis of an increasing 
charge-lattice interaction.
Another important observation supporting the presence of polaronic
carrier, and therefore an intermediate-strong 
local charge-lattice interaction, is 
the presence of a polaronic peak in the MIR band of doped $Ni$ compound 
\cite{Calvani}\cite{Katsufuji}.

The aim of this paper is to show how non-perturbative 
results obtained in the framework of the Dynamical Mean Field Theory 
(DMFT) can be helpful to understand the low energy behavior of 
the optical conductivity in $Ni$ perovskites.
We shall first summarize the results of our theory of the charge ordered 
(CO) state, then we will show a calculation of the optical conductivity
which is in qualitative agreement with experimental observation in
charge ordered $Ni$ perovskites.
In the weak coupling case the CO transition is the well known Charge Density
Wave instability of the Fermi liquid. 
While in strong coupling there have been several attempts to understand the 
CO ordered state based on mean field approach on strong coupling 
effective Hamiltonians \cite{Micnas} and studies of the ground state
\cite{Aubry-Quemerais}.

To discuss the CO state in a non perturbative 
fashion we consider the simplest model of local electron lattice interaction 
i.e. the Holstein molecular crystal model
\cite{Holstein} 

\begin{equation}
H = \sum_i\frac{P^2_i}{2 M}+\frac{1}{2}kX^2_i- \frac{t}{2\sqrt{z}}\sum_{ij}
c_{i}^\dagger  c_{j} + g \sum_{i} (c_{i}^\dagger c_{i}-n) X_i
\label{Holstein-model}
\end{equation}

where $c_{i}^\dagger (c_{i})$ creates
(destroys) an electron at site $i$, and $X_i$, $P_i$
are the local oscillators displacements and momentum.
Electrons and phonons are coupled via the density fluctuations
($n$ being the average electron density). 
The electrons move on a bipartite lattice of connectivity $z$
and have a band of half-bandwidth $t$.

The main approximation we consider is 
the adiabatic approximation which can be obtained in the limit 
$M\rightarrow \infty$. In this limit 
we neglect the first term in eq. (\ref{Holstein-model})
therefore $X_i$ are constant of motion and can be replaced by $c$-numbers.
This approximation turns out to be valid if the following two conditions 
hold

i) as far as thermodynamic properties are concerned  
temperatures must be greater than the typical phonon energy scale 
$\omega_0=\sqrt{k/M}$\cite{Millis}

ii) as far as spectral properties are concerned 
energies must be greater than the typical phonon energy scale 
$\omega_0$

On the other hand  adiabatic limit allows to solve the model 
with a little amount of numerics
giving the spectral 
properties of electrons in {\it real frequencies}
and statistical properties of the lattice. 
We consider also {{\it spinless electron} to account for a polaronic 
rather than a 
bipolaronic ground state at large couplings.
This restriction even if at a very rough level, mimics the action of
an on site Coulomb repulsion. 

We will apply the machinery of the Dynamical Mean
Field Theory which is the exact solution of local-type interaction
on an infinite coordination lattice $z\rightarrow \infty$ 
(infinite dimensions) \cite{Kotliar}.
To have a non trivial limit a scaling of the
hopping, as in eq. (\ref{Holstein-model}), $t$ 
with the number of neighbors is required.
The DMFT approach maps the problem of locally interacting fermions 
on a lattice
into a single site equivalent problem \cite{Kotliar}.
A detailed study of the Holstein model based on the Montecarlo solution
of the single site problem 
has been first carried out in ref. 
\cite{FJS}. 
This analysis has been extended to the spectral properties of the normal 
state in ref. \cite{Millis} by using the adiabatic limit 
to obtain an  
analytical solution of the single site model.
We extend this analytical approach to the study of the charge ordered 
(CO) state.
We consider here alternate charge ordering in two 
interpenetrating sublattices
$A$ and $B$.
The quantity to be determined self-consistently is the
Lattice Polarization Distribution Function 
(LPDF) $P(X)$. 
Different regimes are related to qualitative changes in  the shape
of $P(X)$.  
Our main results can be summarized by the self-consistent equations 
which determines the LPDF and the local electron Green function
in each sublattice

\beq
P_{A,B}(X) \propto e^{-k X^2/2}
\pi_n (i\omega_n - \frac{t^2}{4} G_{B,A}(i\omega_n)-gX_{A,B})
\label{LPDF}
\eeq

\beq
G_{A,B}(i\omega_n)=\int dX P_{A,B}(X) 
\frac{1}{i\omega_n-\frac{t^2}{4}G_{B,A}(i\omega_n)-gX}
\label{Green}
\eeq

Equations (\ref{LPDF},\ref{Green}) are obtained in the 
simple but non trivial case of 
Bethe lattice
of bandwidth $2t$.
From eq. (\ref{Green}) we see that the Green function
is that of a particle propagating in
a randomly distorted sublattice and sublattice $A$ is coupled to $B$
and vice versa.
The real frequency representation of the 
(retarded) Green function is
simply obtained substituting 
$i\omega_n\rightarrow\omega+i0^+$ in eq. (\ref{Green}), therefore
the adiabatic limit allows to obtain the spectral properties
in real frequencies.

\vskip 30pt
{\large {\bf 2. RESULTS}}
\vskip 15pt
We have solved the self-consistent scheme introduced in the previous section 
by numerical iteration procedure. We consider the 
spinless electron half filled case i.e. one electron each two sites ($n=0.5$). 
We start with an ansatz for the sublattice 
Green function then we get the function $P$ at discrete points trough 
eq. (\ref{LPDF}) and trough 
a numerical integration (eq. (\ref{Green}) we obtain the new $G$.

In the adiabatic limit only one relevant coupling
parameter  measures the electron-lattice interaction
$\lambda=g^2/2kt$. 
It can be expressed as  the ratio of self-trapping energy
(polaron energy $\epsilon_p=g^2/2k$) to the electron kinetic 
energy energy ($t$).
A crossover from strong to weak coupling behavior is expected around 
$\lambda\simeq 1$ \cite{summa-polaronica,piovra}.
These expectations are confirmed by the 
the phase diagram at half filling shown in fig. \ref{phase-diag}. 
The continuous curve represents the CO 
critical temperature
as a function of the coupling strength.
The dashed line marks the normal to polaron crossover
in the normal phase\cite{Millis} and the crossover from 
weak coupling CO (A) to  
strong coupling CO (B).
In both normal and ordered state a crossover line separates
the monomodal and the bimodal behavior of the LPDF, in the ordered state
bimodality appears in the sublattice LPDF.
Results are summarized in fig. \ref{fig-LPDF}.
The typical weak coupling behavior of LPDF across the transition 
temperature is shown in fig. \ref{fig-LPDF} a).
Upon decreasing the temperature an uniform polarization of a given 
sublattice arises. 
The other sublattice, whose LPDF is not shown in the figure, 
develops an opposite polarization so that the net total polarization is zero
as should be for the hamiltonian eq. (\ref{Holstein-model})
which couples deformation and density fluctuations.  
The variance decreases with decreasing temperature.

Moving to polaronic non ordered state 
the LPDF is clearly bimodal and symmetric at half filling
as it is seen from fig. \ref{fig-LPDF} c) (dashed lines).
Upon decreasing the temperature below $T_c$ the sublattice LPDF 
unbalances in favor of a net sublattice polarization
but still remain bimodal.
The weight of the secondary peak decreases
by a further decreasing of temperature.
We explain this secondary peak as due to
temperature activated {\it defects} in the CO 
ordered state \cite{Aubry-Quemerais}.

The bimodal behavior of LPDF which is clear at very large coupling becomes
less pronounced at intermediate coupling (see fig. \ref{fig-LPDF} b)).
In this region in both non ordered and ordered state near $T_c$ we observe that
a non negligible amount of sites are nearly undistorted. 
In this intermediate region of the coupling $0.69>\lambda>0.59$ 
we observe a strong non Gaussian behavior of LPDF. 
Even if a secondary peak is present in the normal phase 
it is not well pronounced in this region. As the temperature is lowered below 
$T_c$ it may happen that this secondary peak is washed out in the ordered 
phase but a pronounced shoulder remains in the LPDF. It eventually develops
again a secondary peak upon a further decrease in temperature.

The Optical Conductivity is obtained once the local Green functions
of the two sublattices are known by a generalization of the Kubo formula.

Details of the calculation will be presented elsewhere.
We show in fig. \ref{fig-OC} the results obtained
at half filling for three different values of the coupling constant
characteristic of small, intermediate and large couplings.
We see that at small coupling (fig. \ref{fig-OC} a))
the optical conductivity of the normal state shows a peak 
at $\omega\simeq 0$ reminiscent of the classical Drude behavior.
This peak is shifted in the CO state at $\omega=2 \Delta$ where $\Delta$
is the CDW gap. 
The position of the peak depends on 
temperature and is shifted toward higher energies as the 
temperature decreases following the enhancing of the order 
parameter.
As the coupling is increased 
(fig.  \ref{fig-OC} b)) a peak at $\omega>0$ {\it is still present} also 
at $T>T_c$ indicating the presence of polarons in the disordered
phase. We notice however a shift of the spectral weight from low
to high energies as the temperature is decreased.
This effect 
is less evident at stronger couplings 
(\ref{fig-OC} c)) because in this case almost all sites are 
polarized (see. fig. \ref{fig-LPDF} c)) and consequently we have 
no appreciable spectral weight at low energy. 
In any case whenever polarons are present in the non ordered state 
a shift toward larger energies of the spectral weight and a
change in the temperature dependence of the amplitude of the peak
is observed in the ordered state.

A strong coupling approximation, equivalent to 
Reik's approximation\cite{Reik} at high 
temperatures \cite{Emin}, 
gives the interpretation
of this spectral weight shift in terms of the number of 
defects in the CO state.
In this approximation the ratio of the optical 
conductivity in the CO state to the same quantity
in the normal state is
\beq
\frac{\sigma_{CO}(\omega)}{\sigma_{norm}(\omega)} = 1-(1/2-n_d)^2 F(T,\omega)
\label{OC-sc}.
\eeq
In eq. (\ref{OC-sc}) $n_d$ is number of defects 
in the charge ordered state which depends on $T-T_c$ whereas 
$F(T,\omega)$ is a function of $T$ only
\beq
F(T,\omega)=\sqrt{\beta/2\pi\epsilon_p} 
\int d\nu\;e^{-\nu^2/2\epsilon_p}\times\nonumber\\
tanh(\frac{\beta(\nu-\omega/2)}{2})
tanh(\frac{\beta(\nu+\omega/2)}{2})
\eeq.
From this formula it is easy to obtain a spectral weight shift
proportional to $(1-2n_d)^2$ from low to high energy.

It is worth to note that the shift of the spectral weight from low to high
energy is a phenomenon actually observed in the $Ni$ oxides
together with an anomalous ratio $2\Delta(0)/T_c=13$ \cite{Katsufuji}.
Both observations are consistent with an intermediate coupling
charge ordering of polarons.
\vskip 30pt
{\large {\bf 3. CONCLUSIONS}}
\vskip 15pt
The  crossover from weak to strong coupling 
CO have been studied in details by introducing a LPDF.
The qualitative change from monomodal to bimodal behaviour 
of this function is interpreted as existence of defects in the ordered phase.
In terms of the defects activation we obtain a qualitative understanding of 
the shift from low to high frequency 
in the spectral weight observed below the CO transition 
in $Ni$ perovskites. 

\section*{Acknowledgements} 

Authors acknowledge useful discussions with D. Feinberg 
and P. Calvani.

%
%
\newpage
{\centerline{\bf Figure caption}}

\begin{itemize}

\item[Fig. 1]{The phase diagram at half-filling. Dots are the 
data obtained from numerical solution of 
eqs. (\ref{LPDF},\ref{Green}), dashed line marks
the separation from unimodal LPDF to bimodal LPDF in both normal and CO
state.}

\item[Fig. 2]{LPDF of the sublattice $A$ 
above (dashed) and below (solid $T_c$)
for a) $\lambda=0.4$ b) $\lambda=1$ c) $\lambda=2$. 
Temperature ranges are 
a) from $T_{in}=5\times10^{-3}$ to $T_{fin}=3\times10^{-2}$ with
temperature step $\Delta T=5\times10^{-3}$, 
b) $T_{in}=2\times10^{-2}$, $T_{fin}=6\times10^{-2}$, 
$\Delta T=5\times10^{-3}$,   
c) $T_{in}=2\times10^{-2}$, $T_{fin}=5\times10^{-2}$, 
$\Delta T=2.5\times10^{-3}$. Temperature is in units of 
the half-bandwidth $t$.}

\item[Fig. 3]{Optical conductivity 
above (dashed) and below (solid $T_c$)
for a) $\lambda=0.4$ b) $\lambda=1$ c) $\lambda=2$. Temperature 
ranges are the same as in fig. \ref{fig-LPDF}}   
\protect\label{fig-OC}
\end{itemize}

\newpage
\begin{figure}[htbp]
\centering\epsfig{file=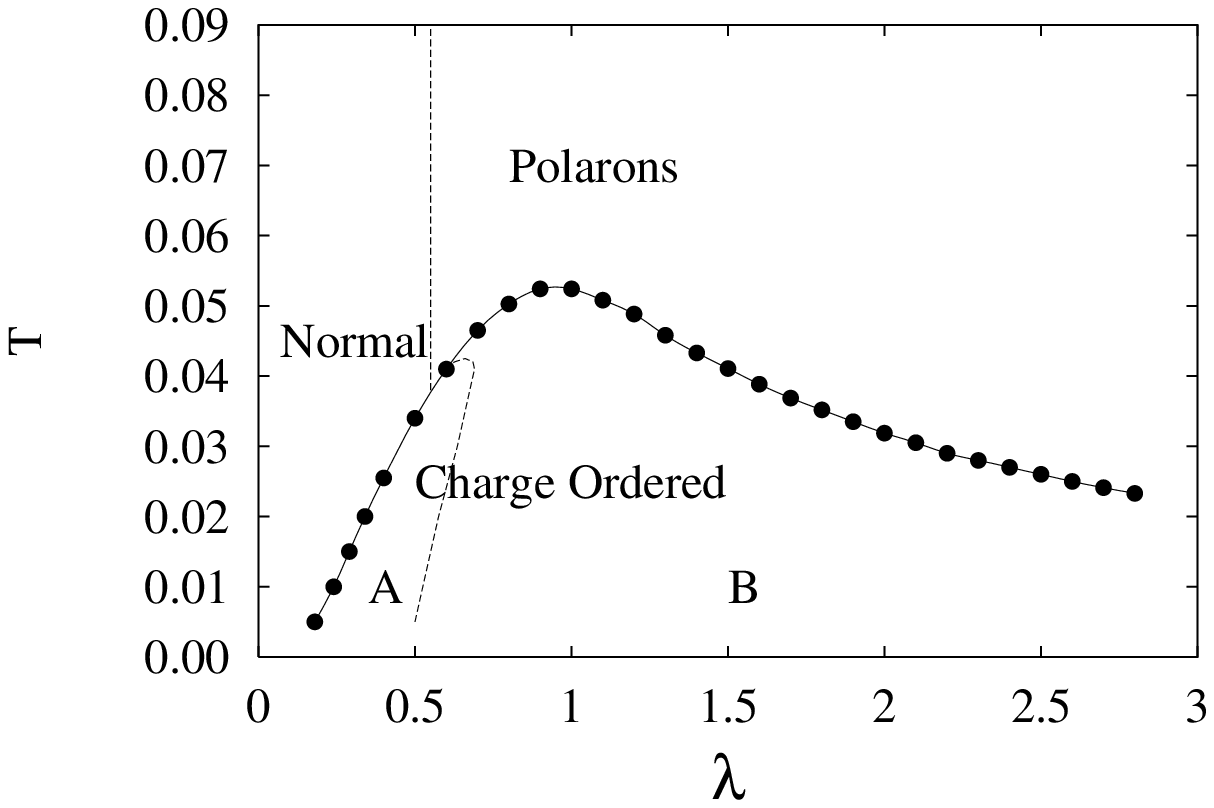,width=1.0\linewidth}  
\protect\label{phase-diag}
\end{figure}
\newpage
\begin{figure}[htbp]
\centering\epsfig{file=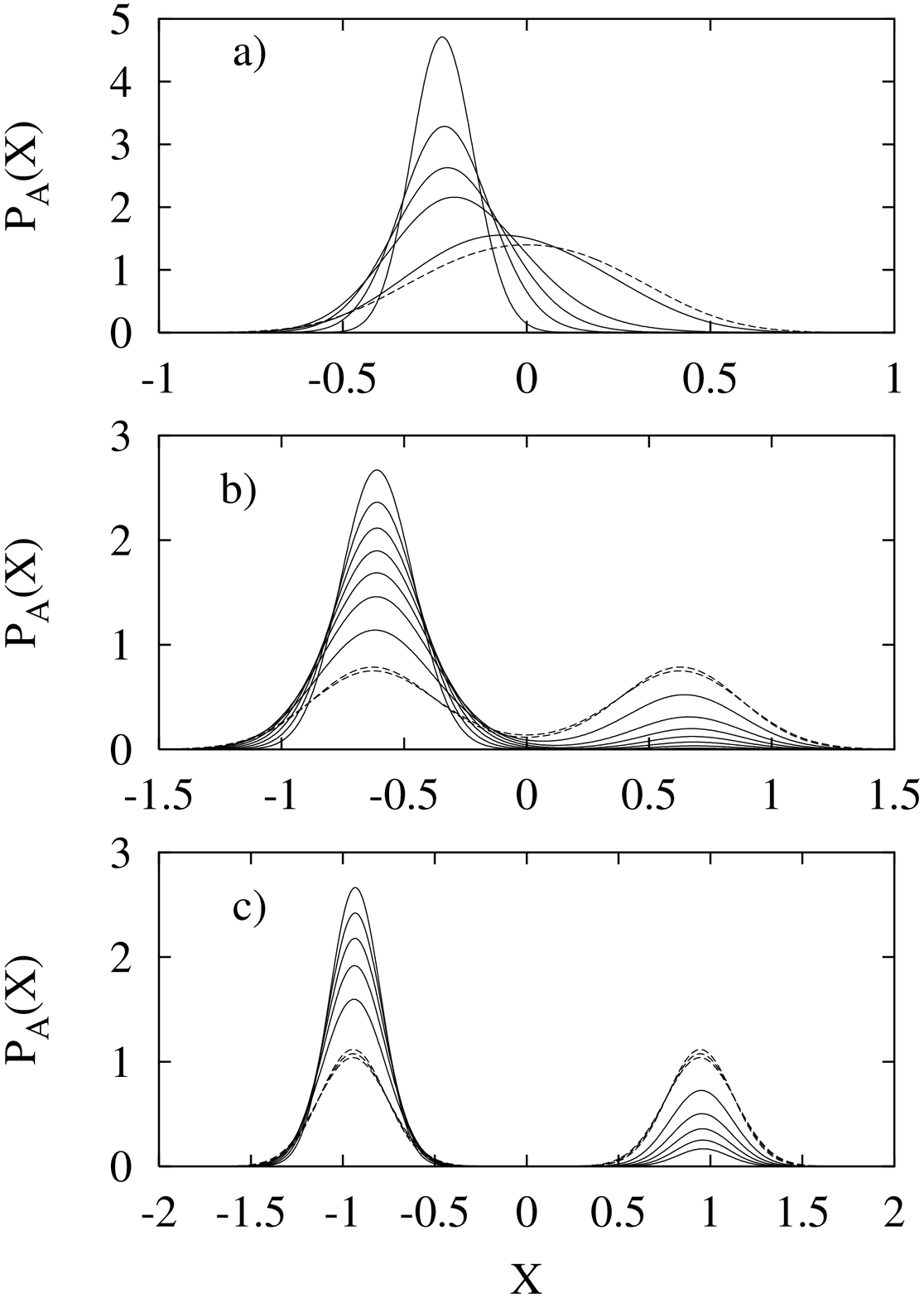,width=1.0\linewidth}
\protect\label{fig-LPDF}
\end{figure}
\newpage
\begin{figure}[htbp]
\centering\epsfig{file=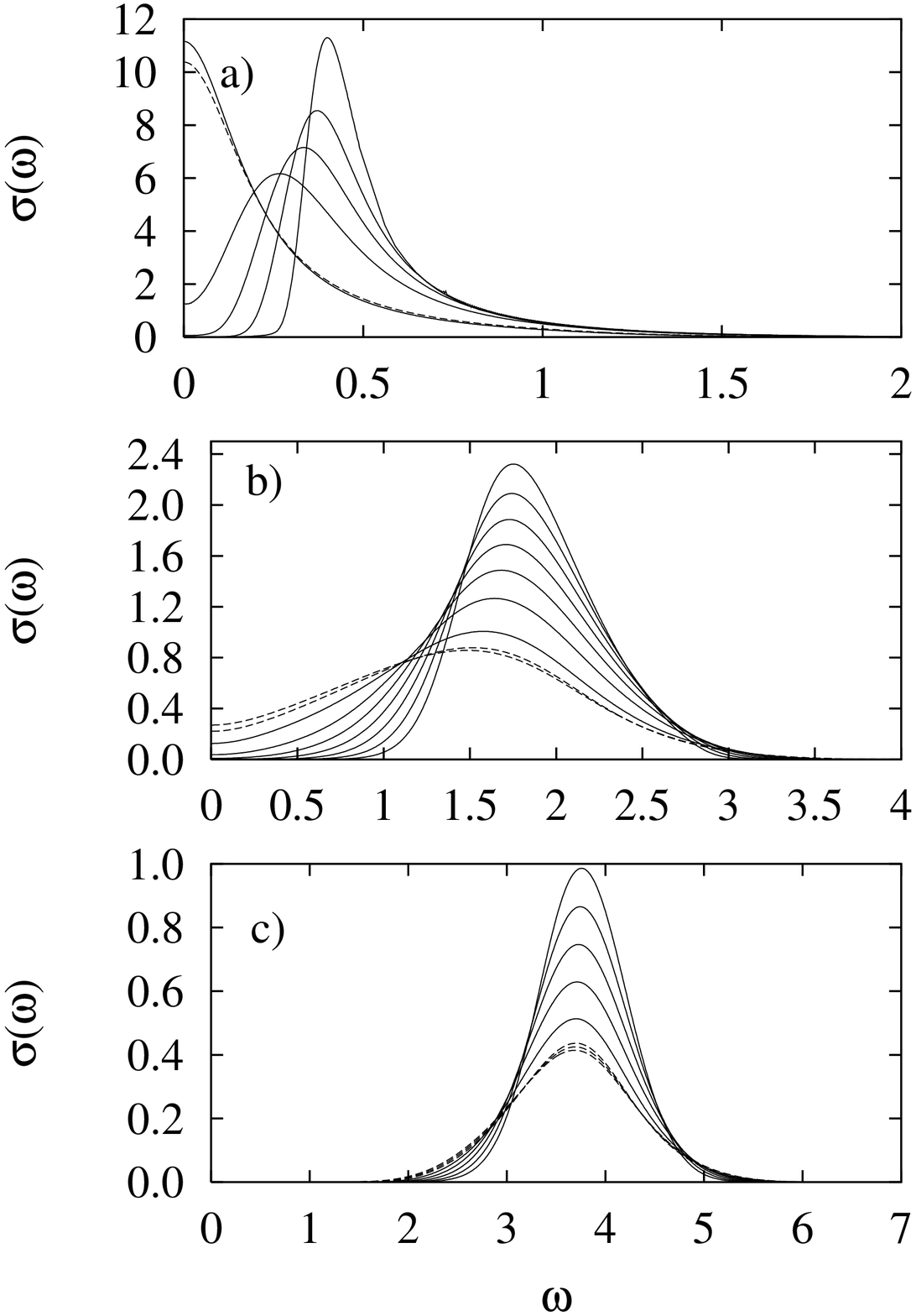,width=1.0\linewidth}  
\end{figure}

\begin{thebibliography}{99}
\bibitem{LaNd} J.M.Tranquada, B.J.Sternlieb,J.D.Axe, Y.Nakamura and S.Uchida
Nat. {\bf 375}, 563 (1995)
\bibitem{LASCO} J.M.Tranquada preprint {\tt cond-mat/9802043}
\bibitem{BISCO} A.Bianconi M.Lusignoli, N.L.Saini,
P.Bordet,A.Kvick and P.G.Radaelli, Phys. Rev. B {\bf 54}, 4310 (1996)\\
A.Bianconi, N.L.Saini, T.Rossetti, A.Lanzara, A.Perali M.Missori,H.Oyanagi,
H.Yamaguchi and Y.Nishitara, D.H.Ha, Phys. Rev. B {\bf 54}, 12018 (1996)
\bibitem{Ni} P.Worchner J.M.Tranquada,D.J.Buttrey and V.Sachan
Phys. Rev. B {\bf 57}, 1066 (1998) 
\bibitem{Mn} C.H. Chen and S-W.Cheong, Phys. Rev. Lett {\bf 76}, 4042 (1996)\\
P.G.Radaelli, D.E.Cox. M.Marezio and S-W. Cheong. Phys. Rev B {\bf 55}, 
3015 (1997)
\bibitem{Calvani}
P.Calvani, P.Dore, S.Lupi, A.Paolone,P.Maselli, P.Giura
B.Ruzicka, S-W. Cheong and W. Sadowsky, J.Supercond. {\bf 10} (1997) 193\\ 
P.Calvani,  A.Paolone, P.Dore, S.Lupi, P.Maselli, P.C. Medaglia
and S-W. Cheong, Phys Rev. B {\bf 54}  R9692 (1996)
\bibitem{Katsufuji} T. Katsufuji, T.Tanabe, T. Ishigawa, Y. Fukuda, T. Arima
and Y. Tokura, Phys. Rev. B  {\bf 54}  R14230 (1996)
\bibitem{Micnas} R. Micnas J. Ranninger and S. Robaszkiewicz
Rev. Mod. Phys. {\bf 62} 113, (1990)
\bibitem{Aubry-Quemerais} S.Aubry and P. Quemerais "Breaking of analiticity
in charge-density wave systems physical interpretation and consequences"
Low Dimensional Electronic Properties of Molybdenium Bronzes
and Oxides, C. Schlenker ed. (Kluwer Academic Publishers,  Dordrecht,
Boston, London, 1987)
\bibitem{Holstein} T.Holstein, Ann.Phys.
{\bf 8}, 325 (1959), {\it ibid.} 343  (1959)
\bibitem{Millis}
A.J.Millis, R. Mueller and B. I. Shraiman, Phys. Rev. B {\bf 54} ,5389,
(1996)
\bibitem{summa-polaronica} S.Ciuchi F.de Pasquale S. Fratini and D.Feinberg
Phys. Rev. B 56, 4418 (1997)
\bibitem{piovra} M.Capone, S.Ciuchi and C.Grimaldi, 
Europhys. Lett. {\bf 42}, 523 (1998)
\bibitem{Kotliar} A.Georges, G.Kotliar,
W.Krauth and M.J.Rozenberg,  Rev.Mod.Phys. {\bf 68},13  (1996) 
\bibitem{FJS} J.K.Freericks,M.Jarrel,D.J.Scalapino Phys. Rev. B {\bf 48}, 
6302 (1993)
\bibitem{Reik} H.G. Reik Z.Phys B {\bf 203}, 346 (1967)
\bibitem{Emin} D. Emin Phys. Rev. B {\bf 48}, 13692 (1993)

\end{thebibliography}
\end{document}